\documentclass{aa}  

\usepackage{amsmath}
\usepackage{graphicx}
\usepackage{float} 
\usepackage{txfonts}
\usepackage{stfloats}
\usepackage{placeins}

\usepackage[colorlinks, citecolor=blue, linkcolor=blue]{hyperref}

\usepackage{subcaption} 
\usepackage[utf8]{inputenc}
\begin{document}

   \title{From \textit{Gaia} to \textit{GaiaNIR}: \\I. Probing dark matter halos in globular clusters}

   \author{I. Henum\inst{1}
          \and
          D. Hobbs\inst{1}
          \and
          \'O. Jim\'enez-Arranz\inst{1}
          \and
          P. J. McMillan\inst{2}
          \and
          R.~P. Church\inst{1}
          }

   \institute{Lund Observatory, Division of Astrophysics, Department of Physics, Lund University, Box 118, SE-22100, Lund, Sweden
         \and
             School of Physics \& Astronomy, University of Leicester, University Road, Leicester LE1 7RH, UK
             \\
             }

   \date{Received <date> / Accepted <date>}

\abstract{
The proposed \textit{GaiaNIR} mission would extend \textit{Gaia}'s astrometric capabilities into the near-infrared, improving astrometric precision and enabling observations in heavily dust-obscured regions. In this work, we investigate the impact of \textit{GaiaNIR} on the detectability of dark matter halos in globular clusters by comparing its performance with that of \textit{Gaia}. Expected observations from future \textit{Gaia} data releases and \textit{GaiaNIR} are modeled for a globular cluster with properties similar to M4. The cluster is simulated with a range of dark matter halo sizes and varying levels of extinction, allowing a direct assessment of each mission’s ability to detect and distinguish dark matter halos across different extinction conditions. To support this comparison, the relationship between the \textit{Gaia} $G$ band and several near-infrared bands is examined. We find that \textit{Gaia} can resolve extended dark matter halos in low-extinction conditions, but its performance degrades significantly as extinction increases. In contrast, \textit{GaiaNIR} reduces statistical uncertainties and retains sensitivity to extended halos even with strong extinction. These results indicate that \textit{GaiaNIR} would both strengthen constraints on kinematic signatures accessible to \textit{Gaia} and enable the study of clusters in heavily obscured regions that are currently beyond \textit{Gaia}'s reach.}


   \maketitle

\section{Introduction}

Globular clusters are dense, gravitationally bound stellar systems, containing $\sim 10^4- 10^7$ stars. They are among the oldest stellar populations in the Milky Way (MW) and therefore key tracers of the Galaxy's formation and evolutionary history \citep[e.g.][]{GC1, GC2}. Most known globular clusters reside in the Galactic halo, though significant populations are also found in the Galactic bulge and disk \citep[e.g.][]{Gaia_2018b}. 

The formation of globular clusters remains debated, with proposed scenarios ranging from them representing the massive end of the normal star cluster population \citep[e.g.][]{HM_tail} to formation under special conditions, such as galaxy mergers \citep[e.g.][]{1992_mergers}, dense gas flows \citep[e.g.][]{gas}, low-metallicity environments \citep[e.g.][]{low_metallicity}, disk instabilities \citep[e.g.][]{Kravtsov_2005}, or within primordial dark matter halos \citep[e.g.,][]{1984_DM_formation, 2005_DM_formation, Vitral_2022, Taylor_2025}. The latter scenario naturally raises the question of whether globular clusters retain dark matter today. While traditionally assumed to be baryon-dominated systems \citep[e.g.][]{Reynoso_Cordova_2022, Taylor_2025}, the detection of even a small dark matter component would place strong constraints on their formation history and on the nature of dark matter on sub-galactic scales. 

The European Space Agency's (ESA) \textit{Gaia} mission \citep{gaia16, gaia2018, gaia23} has revolutionized globular cluster studies by providing high-precision astrometric and photometric measurements at visible wavelengths, enabling the determination of accurate cluster memberships, distances, systemic motions, and internal kinematics for a large fraction of the MW globular cluster population \citep[e.g.][]{Baumgardt_2018, Vasiliev_2021}. However, the observations of globular clusters can be significantly affected by interstellar extinction, depending on their Galactic position. Halo clusters are generally less obscured, whereas clusters located in the Galactic bulge and inner disk experience high levels of extinction \citep[e.g.][]{dharmawardena2024allskythreedimensionaldustdensity, barbillon20253dextinctionmapsmilky}. In these regions, dust severely limits \textit{Gaia}'s ability to accurately measure stellar positions and magnitudes, reducing both the completeness and precision of observations \citep[e.g.][]{Gaia_extinction_2019, Gaia_extinction_2023}. Yet these regions remain crucial for understanding star formation and the dynamical evolution of the inner MW.

The proposed \textit{GaiaNIR} mission \citep[][Hobbs et al. in prep.]{2021ExA....51..783H} aims to overcome this limitation by extending astrometric observations into the near-infrared (NIR), where extinction is substantially reduced. This will both provide more precise data for halo clusters already observed by \textit{Gaia}, and enable studies of globular clusters in dusty regions that are currently poorly accessible with \textit{Gaia} alone. By providing more precise and complete astrometric measurements, \textit{GaiaNIR} will place significantly stronger constraints on the internal dynamics of globular clusters, thereby enhancing our ability to measure the possible presence of dark matter. 

In this work, we assess how well both  \textit{Gaia} and \textit{GaiaNIR} can distinguish dark matter halos of different sizes around globular clusters under varying extinction conditions. To do so, we construct a model globular cluster with properties similar to those of Messier 4 (M4), which is the nearest globular cluster at $\sim 1.8~\mathrm{kpc}$ and therefore a well-studied system by \textit{Gaia} \citep[e.g.][]{H_nault_Brunet_2018, Vitral_2023}. We use this M4-like cluster model as a baseline and modify it by embedding it within dark matter halos of different scale radii and by applying varying levels of interstellar extinction. This allows us to assess how dark matter halos influence the cluster’s observable kinematics under different extinction conditions, and how these differ between \textit{Gaia} and \textit{GaiaNIR}.

This paper is organized as follows: In Sect. \ref{Model cluster construction}, we model the cluster, and assign the necessary spectral and photometric properties to its stars to enable mock \textit{Gaia} and \textit{GaiaNIR} observations. We also embed the cluster in dark matter halos of different sizes, so we can study their effect on its internal kinematics. In Sect. \ref{Simulation of observational effects} we simulate the expected astrometric performance of \textit{Gaia} and \textit{GaiaNIR}, and apply interstellar extinction to the model clusters. We also investigate visible–infrared photometric transformations by comparing the \textit{Gaia} $G$ band with NIR passbands. In Sect. \ref{Gaia and GaiaNIR mock observations} we use the simulated performance of \textit{Gaia} and \textit{GaiaNIR} to recover the observed velocity dispersion profiles of the different cluster models, under varying levels of interstellar extinction. In Sect. \ref{Results}, we present the resulting velocity dispersion profiles, and in Sect. \ref{Discussion}, we interpret these results, discuss their limitations, and outline directions for future work. Finally, in Sect. \ref{conlusions}, we summarize our findings.

\section{Model cluster construction}
\label{Model cluster construction}

We start by constructing a model globular cluster in Sect. \ref{Cluster generation and coordinate systems}, with properties and placement in the Galaxy comparable to those of M4. We then assign spectral energy distribution templates and derive the \textit{Gaia} $G$ magnitudes for each star in the model cluster in Sect. \ref{Spectral template assignment}, as this is required for later \textit{Gaia} and \textit{GaiaNIR} mock observations. In Sect. \ref{Dark Matter Halo Modeling} we modify the cluster dynamics by embedding the system in dark matter halos of different sizes. In Sect. \ref{Velocity dispersion modeling and velocity assignment}, we derive the velocity dispersion profiles of the different dark matter halo models and use them to assign new velocities to the cluster stars.

\subsection{Cluster generation and coordinate transformation}
\label{Cluster generation and coordinate systems}

As a first step, we model a globular cluster using the open-source tool \texttt{McLuster} \citep{McLuster}, which is designed to generate artificial star clusters. The code outputs three-dimensional positions and velocities for each star, along with stellar parameters such as mass, luminosity, effective temperature, and photometric magnitudes, all evolved to a specified cluster age.

We generate a model with a present day mass of $M = 5 \times 10^4 M_\odot$ following a King density profile \citep{1966-King} with concentration parameter $W_0=7$ and half-mass radius of $r_m = 4~\mathrm{pc}$. The cluster generated has 156~195 stars. We evolve the stars to an age of $t = 12~\mathrm{Gyr}$, representing an old globular cluster, and assign a metallicity of $Z=0.001$, typical of metal-poor globular clusters such as M4 \citep{M4-Metal}. We do not include mass segregation or binaries. We place the model cluster at Galactocentric Cartesian coordinates $(X,Y,Z) = (-6373,~-277,~523)~\mathrm{pc}$, at a distance of $1850~\mathrm{pc}$ from the Sun, corresponding to M4's location in the Galaxy \citep{Goldsbury_2010, PM_mean_cut}. 

The stellar positions produced by \texttt{McLuster}, given relative to the cluster center, are transformed into equatorial coordinates (RA, Dec) in the ICRS using \texttt{Astropy}'s \texttt{SkyCoord} class \citep{AstroPy}, adopting the default Galactocentric parameters of \texttt{v5.2.2}. The model velocities $(v_x, v_y, v_z)$ are likewise converted into observable quantities, namely proper motions and line-of-sight velocities. We subtract the mean proper motion and radial velocity to set the systemic motion of the cluster to zero. The resulting model therefore reproduces M4's on-sky position and internal kinematics, but does not include its motion through the Galaxy.

\subsection{Spectral template and G magnitude assignment}
\label{Spectral template assignment}

To simulate how \textit{Gaia} and \textit{GaiaNIR} observe the model cluster (described later in Sect. \ref{Instrumental performance and astrometric errors}), each star in the model cluster is assigned a template spectrum from the Pickles Atlas \citep{Pickles_Atlas}. The atlas contains 44 templates for main-sequence (V) and giant (III) stars, which are the stellar types most commonly observed by \textit{Gaia}. In \texttt{McLuster}, the parameter \texttt{kw} encodes the stellar evolutionary stage of each star. Following the stellar-type classification of \citet{Hurley_2000}, values of $\texttt{kw}=0-2$ are assigned to main-sequence stellar spectra, while values of $\texttt{kw}=3-6$ are assigned to giant stellar spectra.\footnote{Stars with $\texttt{kw}=2$ correspond to Hertzsprung-gap stars in the \citet{Hurley_2000} classification. In this work they are approximated by main-sequence templates.} Stars with $\texttt{kw} > 6$ correspond to helium stars and stellar remnants and are excluded from the analysis. After this cut, 138~324 stars remain out of the initial 156~195 stars in the model cluster. The appropriate main-sequence or giant template is selected for a star based on effective temperature, a parameter provided both in the Pickles Atlas and in the \texttt{McLuster} output. Each star is matched to the template with the closest effective temperature.

The Pickles spectra are provided in units of $\mathrm{erg~cm^{-2}~s^{-1}~\AA^{-1}}$ and are normalized to $V=0$ in the Vega magnitude system. To obtain absolute fluxes, each template is rescaled using the flux of Vega at $5556~\text{\AA}$, $F_\mathrm{Vega}(5556~\text{\AA}) = 3.46 \times 10^{-9}~\mathrm{erg~cm^{-2}~s^{-1}~\AA^{-1}}$ \citep{Vega}. The spectra are then converted from energy flux to photon flux by dividing by the photon energy at each wavelength, $E_{\gamma}(\lambda) = hc/\lambda$, since the precision of \textit{Gaia} and \textit{GaiaNIR} measurements depend on photon count.

Since \textit{Gaia} measures in the $G$ band, apparent $G$ band magnitudes are derived for all model stars to simulate \textit{Gaia} observations. Using the Johnson–Cousins relations for \textit{Gaia} DR3 \citep{esa_phot_relations}, the apparent $G$ magnitude of each model star is computed from its apparent $V$ magnitude and $B-V$ color index, both provided by \texttt{McLuster} (Eq.~\ref{G_1}). For the Pickles templates, $G$ magnitudes are computed from their $V-I_c$ colors assuming $V=0~\mathrm{mag}$ (Eq.~\ref{G_2}). Details of these transformations are given in Appendix~\ref{Empirical relations}.

The Pickles spectra are then rescaled to match the brightness of the corresponding model star. This is done using the difference between the model star’s apparent $G$ magnitude ($G_1$) and the template $G$ magnitude ($G_2$), applying Pogson’s law: $F_1/F_2 = 10^{-0.4 (G_1 - G_2)}$. This scaling ensures that the adjusted spectrum reproduces the apparent brightness of the model star. The Pickles Atlas templates assume solar metallicity, whereas globular clusters such as M4 have $[\mathrm{Fe/H}] \approx -1$. Although the templates are rescaled to match the metallicity-dependent $V$ magnitudes from \texttt{McLuster}, metallicity-dependent spectral features are not necessarily fully reproduced. We assessed the impact of this mismatch using PARSEC isochrones \citep{Parsec} and found that the resulting photometric differences are negligible.

\subsection{Dark matter halo modeling}
\label{Dark Matter Halo Modeling}

To investigate how \textit{Gaia} and \textit{GaiaNIR} would observe a globular cluster embedded in a dark matter halo, we construct two versions of the model cluster: one containing only stellar mass and one including a dark matter halo. The stellar component is modeled by a King model, whose density distribution is given by:

\begin{equation}
\rho(\Psi) = \rho_1 \left[e^{\Psi/\sigma^2}\operatorname{erf}\!\left(\frac{\sqrt{\Psi}}{\sigma}\right) - \sqrt{\frac{4\Psi}{\pi\sigma^2}}\left(1 + \frac{2\Psi}{3\sigma^2}\right)\right]
\label{eq:king_density}
\end{equation}

where erf is the error function, $\Psi\equiv-\Phi+\Phi_0$ is the relative potential, $\Phi$ is the gravitational potential, and $\Phi_0$ is some constant. Also, $\rho_1$ is the central density, assumed to be $\rho_1 = 2~\mathrm{{M}_{\odot}~pc^{-3}}$, and $\sigma$ is the constant velocity dispersion, assumed to be $\sigma = 3.6~\mathrm{km~s^{-1}}$. These parameters are chosen to reproduce a cluster with mass and size comparable to the \texttt{McLuster} model, described in Sect. \ref{Cluster generation and coordinate systems} and to the observed properties of M4. To determine the density and potential as functions of radius, we solve Poisson's equation in spherical symmetry. We assume that $\Psi(0) = W_0 \sigma^2$, where $W_0=7$ is the central dimensionless potential, and  that $d\Phi/dr = 0$ at $r=0$, due to spherical symmetry. The density ($\rho$), potential ($\Phi$), and mass ($M$) profiles of the King model as function of radius are illustrated in Fig.\ref{fig:King_vs_NFW_Profiles} (left panel).

The dark matter component is modeled with a Navarro-Frenk-White (NFW; \citealt{NFW}) model, whose gravitational potential is given by:

\begin{equation}
\Phi(r) = - 4\pi G\rho_0 r_s^2 \frac{\ln(1 + r/r_s)}{r},
\end{equation}

where $r_s$ is the scale radius, which we vary between $8$, $12$, $16$, and $20~\mathrm{pc}$. $\rho_0$ is the characteristic density, taken to be $\rho_0 = 11~\mathrm{M_\odot pc^{-3}}$, which sets the overall dark matter density and produces dark matter masses comparable to the stellar mass. The density profile of the NFW halo is given by:

\begin{equation}
    \rho(r) = \frac{\rho_0}{\frac{r}{r_s} \left(1 + \frac{r}{r_s} \right)^2}
\label{NFW_density}
\end{equation}

The density ($\rho$), potential ($\Phi$), and mass ($M$) profiles of the NFW model as functions of radius are shown in Figure~\ref{fig:King_vs_NFW_Profiles} (right panel). Above, we have followed the standard King and NFW formalisms as presented in \citet{binney1998galactic}, and a more detailed derivation of the profiles can be found in Appendix \ref{King and NFW profiles}. 

Finally, to incorporate the dark matter halo into the stellar cluster, the total gravitational potential is constructed as the sum of the stellar and dark matter contributions: $\Phi(r) = \Phi_\text{King}(r) + \Phi_\text{NFW}(r)$. Only the stellar component described by the King model is observable, so the dark matter halo is included only in the gravitational potential, which in turn determines the stellar kinematics.

\subsection{Velocity dispersion modeling}
\label{Velocity dispersion modeling and velocity assignment}

The internal velocity dispersion profile describes the statistical spread of stellar velocities and is directly determined by the gravitational potential of the cluster. It therefore provides a primary observable for testing the underlying mass distribution and for identifying possible dark matter components. In our models, we compute the velocity dispersion profiles by solving the spherical Jeans equation under the assumption of velocity isotropy, given by:

\begin{equation}
\frac{d (\rho \sigma_r^2)}{dr}  
+ \rho \frac{d \Phi}{dr} = 0
\label{Jeans}
\end{equation}

where $\rho$ is the stellar density and $\Phi$ is the total gravitational potential. The boundary condition $\sigma_r = 0$ is imposed at the outer radius. For the cluster without dark matter, $\Phi$ includes only the King potential, whereas for the dark-matter models $\Phi$ includes both the King and NFW components.

To enable a direct comparison between models with and without dark matter, we scale all velocity dispersion profiles to match at the cluster center. The shape of the velocity dispersion profile is determined by the stellar density distribution and the gravitational potential through the Jeans equation, and is therefore fixed once these are specified. The main remaining freedom is the overall mass normalization, which shifts the profile vertically without significantly changing its shape. Since most of the cluster mass is concentrated toward the center, this region provides the strongest constraint on the mass scale. As a result, matching the profiles at the center effectively corresponds to a renormalization of the total mass. The remaining differences at larger radii therefore reflect the radial mass contribution of the dark matter halo, rather than differences in the total mass.

\begin{figure}
\centering
\includegraphics[width=\columnwidth]{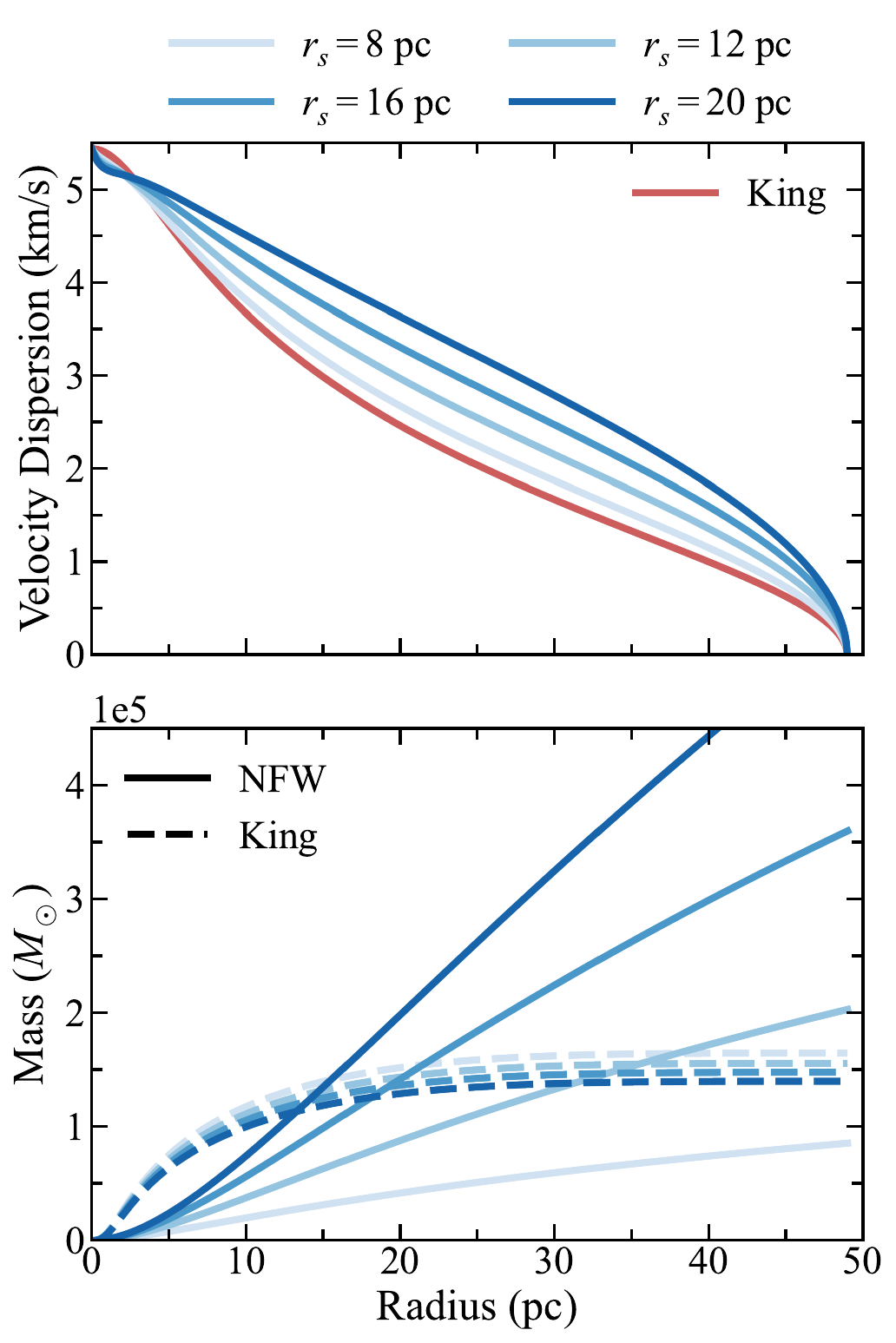}
\caption{Upper panel: Velocity dispersion profiles as a function of radius for the model cluster with a stellar-only component (King model), shown as a solid red line, and for models combining the stellar component with dark matter halos (King+NFW model), shown as solid blue lines. The halo scale radii are varied at $r_s = 8, 12, 16,$ and $20~\mathrm{pc}$.
Lower panel: Relative mass profiles of the stellar (King) and dark matter (NFW) components for the same models shown in the upper panel. The dashed curves represent the stellar mass distribution, while the solid curves show the corresponding dark matter distribution.}
\label{fig:Velocity_and_Mass_Profiles}
\end{figure}

Figure \ref{fig:Velocity_and_Mass_Profiles} (upper panel) shows how the presence of a dark matter halo changes the velocity dispersion profile of the cluster. To explore how the radial extent of the halo influences the velocity dispersion, we vary the NFW scale radius $r_s$ between $r_s= 8, 12, 16$, and $20~\mathrm{pc}$, with larger $r_s$ values shifting the radius at which the dark matter mass dominates over the stellar mass to larger distances. The velocity dispersion of the stellar-only cluster, modeled with the King model, is shown by the lowest curve in red. The velocity dispersion of the stellar cluster combined with dark matter halos of different sizes, modeled with the King+NFW model, is shown by the blue curves. Here, higher $r_s$ values lead to an increase in the velocity dispersion. The magnitude and radial extent of this change depend on the relative contributions of the stellar and dark matter components to the total mass distribution, which are shown in Fig. \ref{fig:Velocity_and_Mass_Profiles} (the lower panel). Here, the solid lines represent the halo mass for different $r_s$ values, and the dashed lines show the corresponding stellar-only component. The differences in the stellar-only profiles arise from the velocity dispersion rescaling applied to match the central values.

To ensure that the model clusters with and without dark matter differ only in their gravitational potential, we keep all stellar positions and intrinsic properties identical to those from the \texttt{McLuster} model described in Sect. \ref{Cluster generation and coordinate systems}, but replace the initial \texttt{McLuster} velocities with new values drawn from a normal distribution based on the derived velocity dispersion profiles. For each star at radius $r$, the three velocity components are drawn from Gaussian distributions: $(v_x, v_y, v_z) \sim \mathcal{N}\big(0, \sigma^2(r)\big)$. Here $\sigma(r)$ is the local velocity dispersion, obtained by solving Eq. \ref{Jeans}. This ensures that the internal kinematics follow either the stellar cluster alone or the cluster embedded in a dark matter halo. Using the updated velocity profiles, we convert the new velocities into observable quantities (proper motions and line-of-sight velocities) following the procedure described in Sect. \ref{Cluster generation and coordinate systems}. We subtract any residual median proper motion, arising due to random sampling, to ensure zero systemic motion. The resulting proper motions thus reflect only the internal kinematics of the cluster.

To study the velocity dispersion profile as a function of radius, we compute the projected distance of each star from the cluster center, defined as: 
\begin{equation}
    R_i = \sqrt{\left([\alpha_i - \bar{\alpha}] \cos \bar{\delta}\right)^2 + (\delta_i - \bar{\delta})^2},
\end{equation}

where  $(\alpha_i, \delta_i)$ are the coordinates of star $i$, and $(\bar{\alpha}, \bar{\delta})$ are the coordinates of the cluster center. The stars are then grouped into 30 logarithmically spaced radial bins between $108$ and $1800~\mathrm{arcsec}$ ($0.03^\circ$ and $0.5^\circ$), ensuring good sampling of both the dense central region and the more diffuse outskirts of the cluster. The bins contain a mean of 2951 stars, with the number per bin ranging from 1166 in the least populated bin to 3741 in the most populated bin.

Within each radial bin, the intrinsic velocity dispersion is measured directly from the proper motions. For the right ascension and declination components, $\mu_{\alpha^*}$ and $\mu_\delta$, we compute the standard deviation of the stars in the bin, yielding $\sigma_{\mu_{\alpha^*}}$ and $\sigma_{\mu_\delta}$, respectively. Assuming isotropy, the two components are combined to obtain the total proper-motion dispersion,

\begin{equation}
\sigma_{\mu} = \sqrt{\frac{\sigma_{\mu_{\alpha^*}}^2 + \sigma_{\mu_\delta}^2}{2}}.
\label{sig_combined}
\end{equation}

Repeating this calculation across all radial bins provides the radial proper-motion dispersion profiles. The dispersions in proper motion units ($\mathrm{mas~yr^{-1}}$), are converted to physical velocities using: $\sigma_v = 4.7405\sigma_\mu d$. Here $\sigma_v$ is in $\mathrm{km~s^{-1}}$, $\sigma_\mu$ is in $\mathrm{mas~yr^{-1}}$, and $d=1.85~\mathrm{kpc}$ is the adopted distance to M4. 

To estimate the statistical uncertainties on the velocity dispersion profiles, we apply a bootstrap resampling procedure within each radial bin. For every bin, we generate $N=1000$ resampled datasets of the stars and recalculate the standard deviation of the proper motions in right ascension and declination. The scatter among these bootstrap realizations provides an estimate of the standard error in each bin. 

\begin{figure}
\centering
\includegraphics[width=\columnwidth]{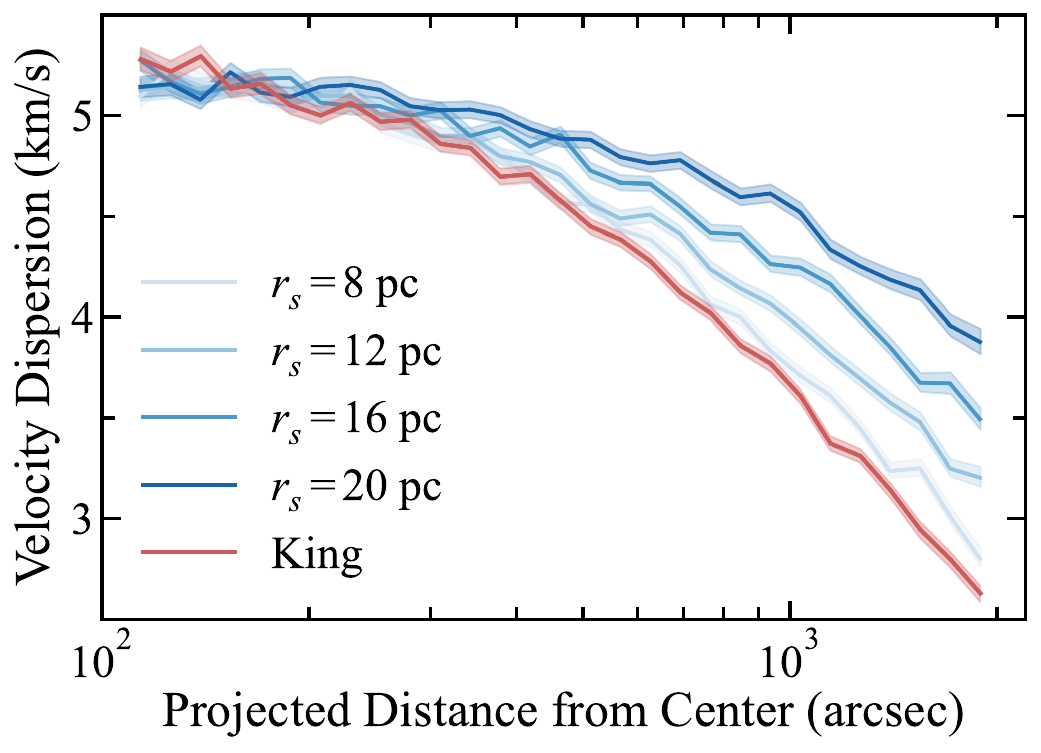}
\caption{Binned velocity dispersion profiles as a function of projected distance from the cluster center, with $1\sigma$ bootstrap errors (no observational errors are added). The red curve shows the stellar-only (King) model, while the blue curves correspond to stellar and dark matter combined (King+NFW) models with halos of different scale radii $r_s = 8, 12, 16,$ and $20~\mathrm{pc}$. }
\label{fig:Velocity_Dispersion_Total_Model}
\end{figure}

In Fig. \ref{fig:Velocity_Dispersion_Total_Model}, the resulting binned velocity dispersions are shown as a function of projected distance from the cluster center for the stellar-only model and for models combining the stellar model with halos of different scale radii, with errorbars derived from the bootstrap procedure. No observational errors has been added to the dispersions yet. As in Fig. \ref{fig:Velocity_and_Mass_Profiles} (upper panel), the velocity dispersion increases with increasing $r_s$ values. In the following, Fig. \ref{fig:Velocity_Dispersion_Total_Model} serves as the intrinsic reference model without observational noise, to which the mock \textit{Gaia} and \textit{GaiaNIR} dispersions are compared.

\section{Simulation of observational effects}
\label{Simulation of observational effects}

Having constructed model clusters with stellar populations and kinematics, both with and without dark matter halos, we now simulate the observational effects, to prepare for mock observations of \textit{Gaia} and \textit{GaiaNIR}. In Sect. \ref{Instrumental performance and astrometric errors} we model the instrumental performance and astrometric uncertainties of both missions, and in Sect. \ref{Interstellar Extinction}, we apply different levels of extinction to the model cluster. We extend the simulations to NIR photometric bands in Sect. \ref{Near-infrared photometric transformations}.

\subsection{Instrumental performance and astrometric errors}
\label{Instrumental performance and astrometric errors}

To investigate how both \textit{Gaia} and \textit{GaiaNIR} will observe globular clusters, we use a simulation program developed by Hobbs et al. (in prep.) that models the instrumental performance of both missions. The program takes as input the apparent $G$ magnitude and the stellar spectrum -- specifically, the photon flux as a function of wavelength drawn from the Pickles Stellar Spectral Flux Library \citep{Pickles_Atlas} -- and computes the predicted end-of-mission parallax uncertainty in $\mu$as.

The simulation integrates the stellar spectrum over the detector quantum efficiency $Q_\lambda$ and instrument transmittance $T_\lambda$ to compute the total number of detected photoelectrons $N$ per transit of the focal plane. The along-scan astrometric uncertainty per transit is then estimated from the photon-noise limit, modified by a readout noise term,
\begin{equation}
\sigma_\eta = \frac{1}{2\pi}\frac{s_2}{\sqrt{N}}\left[\kappa + \frac{(r^2+b),s_0}{N,\Delta\eta}\right]^{1/2},
\end{equation}
where $s_0$ and $s_2$ are characteristic angular parameters of the diffraction-limited PSF, $r$ is the rms readout noise per detector sample, $b$ is the background level (sky background photons and dark current), $\Delta\eta$ is the pixel size, and $\kappa$ is a form factor calibrated from Monte Carlo simulations ($\kappa \approx 0.32$ for bright stars, $\approx 0.25$ at the faint end). The end-of-mission parallax uncertainty is obtained by combining $\sigma_\eta$ over all transits and adding in quadrature a calibration noise floor $\sigma_\mathrm{cal} = 60~\mu$as for \textit{GaiaNIR}:

\begin{equation}
\sigma_{\varpi}  = g_\varpi \left[\frac{\tau_1}{N_i\tau p_{det}(G)}(\sigma^2_{\eta} + \sigma^2_{\rm cal}) \right ] ^{1/2}
\label{eq:eomAccuracy:eomVarpi}
\end{equation}
where $N_i$ is the number of instruments, $\tau=L\Omega/4\pi$ the average total time available per object and instrument, $\tau_1$ is the detector crossing time, and $p_{\rm det}(G)$ is the detection probability as a function of magnitude $G$. It can be noted that $N_i\tau / \tau_1$ is the average total number of detector crossings of the object during the mission, so that $N_i\tau p_{det}(G) / \tau_1$ is the expected number of detected detector crossings, on which the estimation of the astrometric parameters is based. The factor $g_\varpi$ relates the scanning geometry to the determination of the astrometric parameters. It is different for the five parameters (parallax, position at mean epoch in two coordinates, proper motion components in two coordinates) and varies as function of position on the sky, mainly as a function of ecliptic latitude. For {\it Gaia} simulations of the scanning law were used to determine the mean values of $g$ for given mission parameters, and their large-scale variations with ecliptic latitude. 

A critical distinction between \textit{Gaia}'s silicon CCDs and the Avalanche Photo-Diode (APD) detectors proposed for \textit{GaiaNIR} lies in the behaviour of the readout noise term $(r^2+b)$ at faint magnitudes. For \textit{Gaia}'s CCDs, this term dominates at $G\gtrsim 18$, causing the uncertainty to increase exponentially towards the faint limit. The Leonardo APD arrays under development for \textit{GaiaNIR} achieve a sub-electron effective readout noise, which strongly suppresses this term. As a result, the astrometric uncertainty for \textit{GaiaNIR} increases much more slowly with magnitude -- approximately as a power law rather than exponentially -- even for the faintest observed stars. Simulations of a 10 year mission for an M5III star show that at 20.7 magnitude \textit{Gaia}'s CCD gives a parallax uncertainty of $\sim$570~$\mu$as, compared to $\sim$94-45~$\mu$as depending on mission design (5 year or 10 year) for the APD-based detectors, a factor of $\sim$6-13 improvement that propagates directly into all science cases depending on faint stars (Hobbs et al. in prep.).

\addtocounter{footnote}{+1}
\footnotetext{\label{fn:esa}\url{https://www.cosmos.esa.int/web/gaia/science-performance}}
In this work, we simulate the \textit{Gaia} mission performance using time spans of 34 months for DR3, and 5 and 10 years for DR4 and DR5, respectively. Although the actual mission durations for DR4 and DR5 are expected to be approximately 5.5 and 10.5 years, we use 5 and 10 years to ensure consistency with ESA's published performance predictions\footnotemark[\value{footnote}]. The \textit{Gaia} telescope configuration uses an effective aperture $D = 1.45$~m with a focal length of 35~m and an across-scan mirror dimension $H = 0.5$~m.

For \textit{GaiaNIR}, two telescope configurations are implemented: the Medium (M) option with a $D$=1.7~m unobscured single primary mirror, and the Large (L) option with a $D$=3.5~m segmented primary mirror including a 10~cm gap between the two segments. The L configuration is designed to compensate for the loss of angular resolution incurred by observing at longer wavelengths: shifting from \textit{Gaia}'s mean wavelength of $\sim$673~nm to \textit{GaiaNIR}'s $\sim$1550~nm would halve the resolution at fixed aperture, but the 3.5~m primary recovers this, providing an effective baseline $B_\mathrm{eff} = D\sqrt{(1+\varepsilon+\varepsilon^2)/3}$, where the obscuration ratio equivalent to 10~cm $\varepsilon = 0.1/3.5$. We evaluate both the M configuration for 5 and 10 years (M5 and M10) and the L configuration for 5 and 10 years (L5 and L10).

\begin{figure} 
    \centering
    \includegraphics[width=\columnwidth]{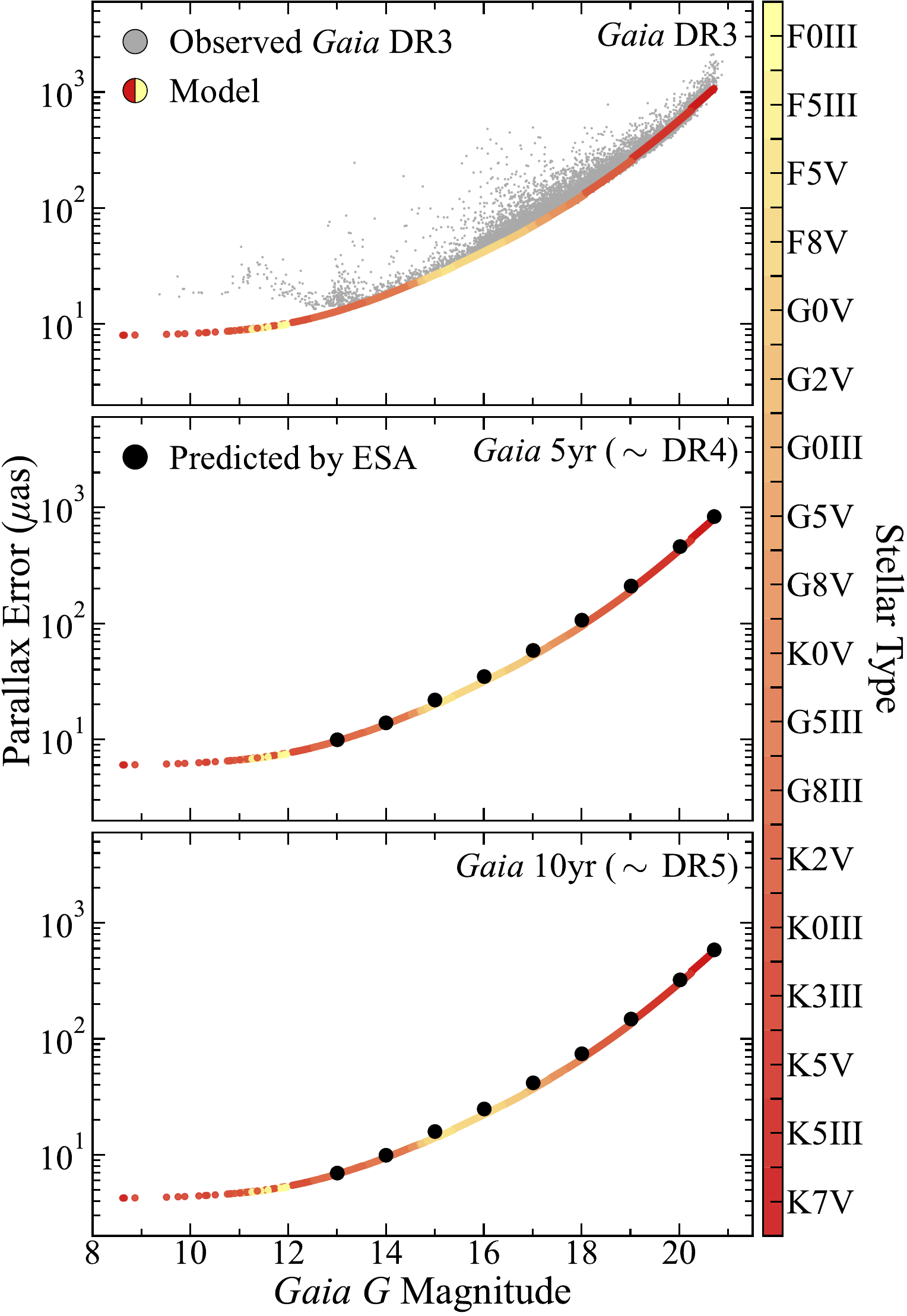}
    \caption{Simulated \textit{Gaia} parallax errors for the model cluster, color-coded by stellar spectral type (earlier hotter types in yellow tones to later cooler types in red tones). Top panel: simulated \textit{Gaia} DR3 parallax errors compared with the actual errors observed for M4 with \textit{Gaia} DR3 in grey dots. Middle panel: simulated DR4 parallax errors compared with ESA's predicted values in black dots. Bottom panel: simulated DR5 parallax errors compared with ESA's predicted values in black dots.}
    \label{fig:Parallax_error_Gaia}
\end{figure}

Figure \ref{fig:Parallax_error_Gaia}  shows the simulated \textit{Gaia} parallax errors for the model cluster as a function of G magnitude, color-coded by stellar type, for DR3, DR4, and DR5. The DR3 simulations are directly compared with the actual parallax errors observed in \textit{Gaia} DR3 for M4 (grey dots; the data selection is described in Sect.~5). The DR4 and DR5 simulations are compared with ESA's predicted performance estimates\footnotemark[\value{footnote}] for 5 and 10 years, respectively (black dots). The good agreement across all magnitudes and mission durations validates the simulation tool before applying it to the \textit{GaiaNIR} configurations. For the model cluster observed with \textit{Gaia}, a cut in $G$ magnitudes is made at $G \leq 20.7$ \citep{gaia16}. After this cut 32~329 stars remain in the cluster. 

\begin{figure}
    \centering
    \includegraphics[
        width=\columnwidth
    ]{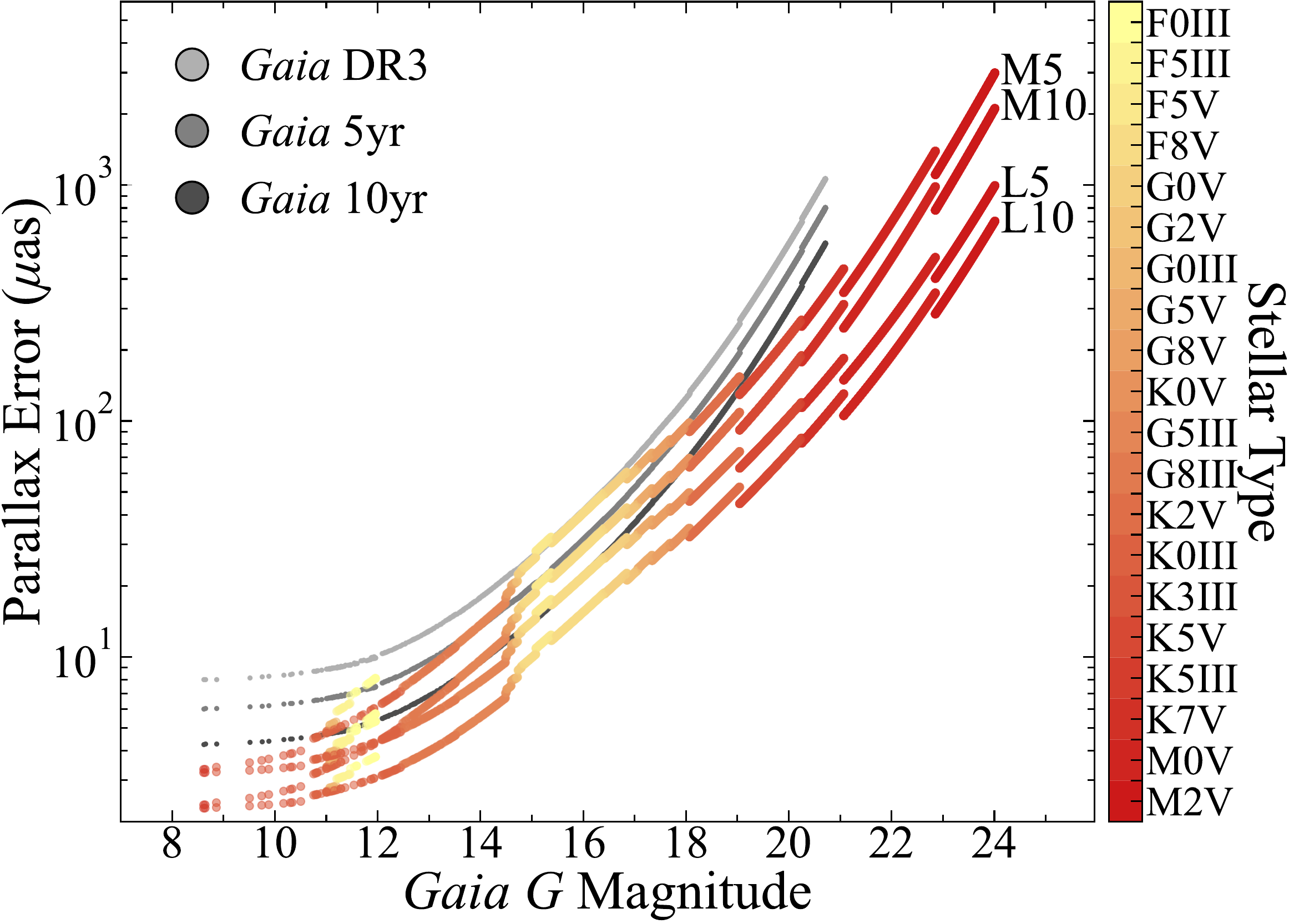}
    \caption{Simulated \textit{GaiaNIR} parallax errors for the model cluster, color-coded by stellar type (earlier hotter types in yellow tones to later cooler types in red tones). The different curves represent the \textit{GaiaNIR} M configurations for 5- and 10-year mission durations, and the \textit{GaiaNIR} L configurations for 5- and 10-year mission durations. The grey curves correspond to the parallax errors of \textit{Gaia} DR3, $\sim$DR4, and $\sim$DR5 from Fig. \ref{fig:Parallax_error_Gaia}.}
    \label{fig:Parallax_error_GaiaNIR}
\end{figure}

Figure \ref{fig:Parallax_error_GaiaNIR} shows the simulated \textit{GaiaNIR} parallax errors for the model cluster, plotted as a function of $G$ magnitude and color-coded by stellar type, for M5, M10, L5 and L10. For comparison purposes, the \textit{Gaia} parallax errors from Fig. \ref{fig:Parallax_error_Gaia} are also plotted in gray. The \textit{GaiaNIR} parallax errors are plotted as a function of \textit{Gaia} $G$ magnitudes. Although \textit{GaiaNIR} measures stars in the NIR, using $G$ magnitudes as the reference magnitude provides a common scale for both missions, and allows a direct comparison with the \textit{Gaia} performance curves. The curves are not smooth as they are for \textit{Gaia}, as stars with the same $G$ magnitude not necessarily are equally bright in the NIR. Cooler stars emit more of their light in the NIR, while hotter stars emit more in the visible. Therefore, at the same $G$ magnitude, the NIR brightness can differ, which causes small variations in the GaiaNIR error curves. For the model cluster observed with \textit{GaiaNIR}, a cut in $G$ magnitudes is made at $G \leq 24$. As the final survey depth is not yet well constrained, this value is adopted as a representative assumption. After this cut 99~461 stars remain in the cluster. From Fig. \ref{fig:Parallax_error_GaiaNIR}, it can be seen that stars at this magnitude cut are mainly of type M5V. For this stellar type, a $G$ magnitude of 24 corresponds to $\sim 21.6$ in the GaiaNIR $N$ magnitude (how this is obtained is explained in Sect. \ref{Near-infrared photometric transformations}). For warmer stellar types (F and G stars), which emit most of their flux in the visible, \textit{Gaia} can outperform \textit{GaiaNIR}, as \textit{Gaia} observes in the $320-1050~\mathrm{nm}$ range, near the peak of their spectral energy distribution, whereas \textit{GaiaNIR} observes in the $800-2300~\mathrm{nm}$ range, where their flux is lower, resulting in fewer detected photons and reduced astrometric precision. However, in practice, \textit{Gaia} and \textit{GaiaNIR} data would be combined, but this is not considered in this work.

To derive the proper motion errors, $\sigma_{\mu_{\alpha^*}}$ and $\sigma_{\mu_\delta}$, from the parallax errors $\sigma_\varpi$, we use the scaling relations from ESA\footnotemark[\value{footnote}], which depend on the duration of the mission. The relations are given as multiplicative factors applied to the parallax error:

\begin{itemize}
    \item \text{34 months mission:}  
    $\sigma_{\mu_\alpha*} = 1.03 \sigma_\varpi$,  
    $\sigma_{\mu_\delta} = 0.89 \sigma_\varpi$
    
    \item \text{5 year mission:}  
    $\sigma_{\mu_\alpha*} = 0.58 \sigma_\varpi$,  
    $\sigma_{\mu_\delta} = 0.50 \sigma_\varpi$
    
    \item \text{10 year mission:}  
    $\sigma_{\mu_\alpha*} = 0.29 \sigma_\varpi$,  
    $\sigma_{\mu_\delta} = 0.25 \sigma_\varpi$
\end{itemize}

These scaling relations are based on \textit{Gaia} performance estimates for DR3, DR4 and DR5, but are also adopted here to approximate the proper motion errors for \textit{GaiaNIR}. It should be noted that the values are sky-averaged. In reality, the errors depend on the position of the cluster on the sky due to the \textit{Gaia} scanning law \citep{Gaia_Early_Data_Release_3_Lennart}. However, for our cluster model this approximation is sufficient, as we focus on the overall astrometric precision and the relative differences between mission scenarios.

\subsection{Interstellar extinction}
\label{Interstellar Extinction}

To simulate observations of the model cluster through dust, we apply interstellar extinction to both the stellar spectra and the apparent $G$ band magnitudes. This allows us to investigate how \textit{Gaia} and \textit{GaiaNIR} performance varies with Galactic environment. For the spectra, we adopt the fit to Seaton’s extinction law \citep{Seaton} and apply extinction to the spectra as outlined in Appendix \ref{Ap:extinction}. 

In addition to modifying the spectral fluxes, extinction is also applied to the apparent $G$ band magnitudes of each star in the model cluster. This ensures that the photometric properties, as observed by \textit{Gaia} and \textit{GaiaNIR}, are affected by interstellar dust in a manner consistent with the spectra. Given the intrinsic and extinguished fluxes, the extinguished $G$ band magnitude can be computed using Pogson's law.

When adding extinction to the model cluster, we test two different values of the color excess: (1) $E(B-V) = 0.37$, matching the measured reddening of M4 \citep{reddening_M4} and representative of regions of moderate extinction. Applying this extinction to the model cluster observed with \textit{Gaia} decreases the number of stars from 32~329 to 19~173. For \textit{GaiaNIR}, the number of stars in the cluster decreases from 99~462 to 79~292. (2) $E(B-V) = 1.0$, characteristic of regions of high extinction within the Galactic disk, particularly near the midplane \citep{Lallement_2014}, where testing the performance of \textit{GaiaNIR} is of particular interest. Applying this extinction to the model cluster observed with \textit{Gaia} decreases the number of stars from 32~329 to 5~863. For \textit{GaiaNIR}, the number of stars decreases from 99~462 to 31~597. In both cases, the placement of the model cluster is unchanged, and only the extinction is varied.

\subsection{NIR photometric transformations}
\label{Near-infrared photometric transformations}

We have illustrated the simulated \textit{GaiaNIR} results in terms of the $G$ magnitude. But, while \textit{Gaia} operates in the visible passbands, \textit{GaiaNIR} extends into the NIR, making it significantly less affected by extinction. To address the differences between the visible and NIR passbands, we investigate color indices between the \textit{Gaia} $G$ band and NIR bands ($J$, $H$, $K_s$) that approximate \textit{GaiaNIR} observations. The transformations between the \textit{Gaia} $G$ band and NIR passbands are adopted from empirical relations, listed in Appendix \ref{Empirical relations}.

We also use in particular a new $N$ band for GaiaNIR \citep{2021ExA....51..783H}, with magnitude defined as: $m_N = -2.5 \log_{10} \left( N_{e,N} / N_{e,\mathrm{Vega}} \right)$, where the $N$ band magnitude is expressed relative to Vega, whose magnitude is defined as $m_N = 0$. $N_{e,N}$ is the total number of detected photoelectrons for a given star in the $N$ band, and $N_{e,\mathrm{Vega}}$ is the corresponding number for Vega. The number of detected photoelectrons is computed in the simulation code, as: $N_e = A \tau Q_\lambda \int F_\lambda T_\lambda~d\lambda$, where $A$ is the telescope aperture, $\tau$ is the effective exposure time, and $Q_\lambda$ is the detector quantum efficiency. The function $F_\lambda$ is the spectral flux density of the star, and $T_\lambda$ is the transmission function of the passband. The integral therefore gives the total detected flux within the filter band. Finally, the corresponding color index is obtained as $G- N = m_G- m_N$.

\begin{figure*}
    \centering
    \includegraphics[width=\textwidth]{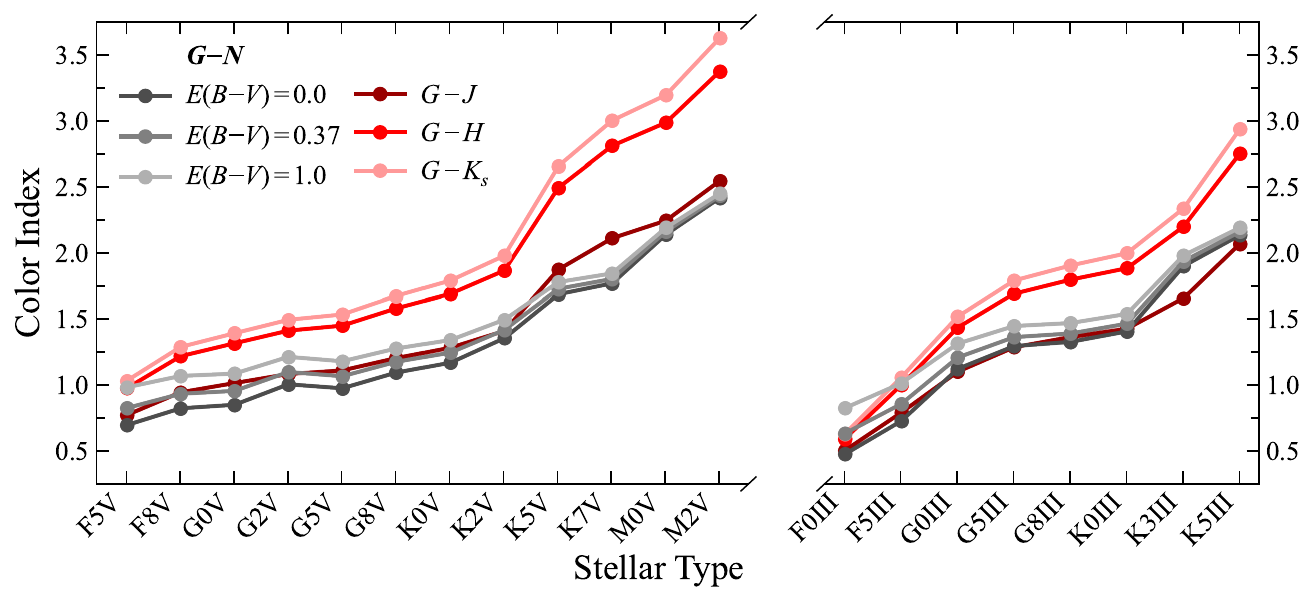}
    \caption{Simulated color indices as a function of stellar type. The \textit{Gaia} $G$ band is combined with NIR passbands ($J$, $H$, $K_s$) and with the \textit{GaiaNIR} $N$ band. Dark grey, grey, and light grey curves show the $G-N$ index without extinction and for two different extinction values, respectively.}
    \label{fig:$G$-N_H_K_J}
\end{figure*}

Figure \ref{fig:$G$-N_H_K_J} shows the visible–NIR color indices as a function of stellar type for both main-sequence stars (left panel) and giants (right panel). The plotted quantities represent the difference between the \textit{Gaia} $G$ magnitude and various NIR magnitudes ($N$, $J$, $H$, and $K_s$). The color indices increase toward cooler stars, as the peak of the emission then shifts to longer wavelengths. The $G-K_s$ index is the largest and the $G-J$ index is the smallest among the ground-based bands, reflecting the larger wavelength separation between the visible $G$ band and the $K_s$ compared to that between $G$ and $J$. The $G-N$ index most closely follows the $G-J$ relation, as the wavelength of the $N$ band is most similar to that of $J$. For the $G-N$ color index, the figure also illustrates the effect of extinction, for reddening values of $E(B-V)=0$, $E(B-V)=0.37$ and $E(B-V)=1$. Increasing extinction shifts the $G-N$ index to higher values, with the magnitude of this effect decreasing toward cooler stars, as their emission is already concentrated at longer wavelengths where extinction is weaker, resulting in a smaller differential color change. 

The $J$, $H$, and $K_s$ bands are ground-based passbands, that are compared with the space-based $N$ and $G$ bands. Notably, space-based NIR observations differ fundamentally from ground-based observations, as the latter are affected by atmospheric absorption, leading to differences in the transmission profiles. Despite these differences, the comparison with ground-based bands provides useful context for interpreting the approximate depth and sensitivity of the $N$ band. To validate our model cluster, we adopted the extinction corresponding to M4 and compared the resulting $H$, $J$, and $K_s$ magnitudes as a function of $G$ with observations, obtained from 2MASS sources cross-matched with \textit{Gaia} DR3, finding good agreement.

\section{\textit{Gaia} and \textit{GaiaNIR} mock observations}
\label{Gaia and GaiaNIR mock observations}

From Sect. \ref{Velocity dispersion modeling and velocity assignment}, we have the model cluster projected onto the plane of the sky, both with and without dark matter halos of different scale radii. Using the proper motion uncertainties of \textit{Gaia} and \textit{GaiaNIR} from Sect. \ref{Instrumental performance and astrometric errors}, we can now simulate how these missions would observe the modeled globular clusters. The observed proper motions are generated by adding measurement noise to the model values, drawing from normal distributions: $\mu_{\mathrm{obs}, \alpha^*} \sim \mathcal{N}(\mu_{\alpha^*}, \sigma_{\mu_{\alpha^*}}^2)$ and $\mu_{\mathrm{obs}, \delta} \sim \mathcal{N}(\mu_{\delta}, \sigma_{\mu_{\delta}}^2)$, where $\mu_{\alpha^*}$ and $\mu_{\delta}$ are the true proper motions from the model, and $\sigma_{\mu_{\alpha^*}}$ and $\sigma_{\mu_{\delta}}$ are the corresponding observational uncertainties simulated for the different missions. This produces mock \textit{Gaia} and \textit{GaiaNIR} data for each cluster model. These mock proper motions, $\mu_{\mathrm{obs},i}$, are treated as the observational data in the following analysis. 

As in Sect. \ref{Velocity dispersion modeling and velocity assignment}, the stars are grouped into radial bins according to their projected distance from the cluster center, allowing the velocity dispersion profile to be measured as a function of radius. Within each radial bin, we estimate the intrinsic velocity dispersion of the cluster using a Maximum Likelihood Estimation (MLE) approach, which treats the data as observations and separates the intrinsic dispersion from the measurement errors. The mock proper motions are assumed to be random realizations drawn from a Gaussian distribution centered on a mean value $v_0$, with a total variance equal to the sum of the intrinsic cluster dispersion, $\sigma_{\mathrm{int}}^2$, and the individual measurement errors, $\sigma_{\mathrm{obs},i}^2$: $\mu_{\mathrm{obs},i} \sim \mathcal{N}(v_0, ~ \sigma_{\mathrm{int}}^2 + \sigma_{{\mathrm{obs}},i}^2)$. Here, $v_0$ represents the systemic proper motion of the cluster in that bin (typically close to zero after subtraction of the bulk motion), while $\sigma_{\mathrm{int}}$ is the intrinsic velocity dispersion to be estimated. 

Assuming all measurements are independent, the likelihood function for a set of $N$ stars in a given radial bin is then expressed as:
\begin{equation}
\mathcal{L} = \prod_{i=1}^{N} \frac{1}{\sqrt{2\pi\left(\sigma_{\rm int}^2 + \sigma_{{\rm obs},i}^2\right)}} 
\exp\!\left[-\frac{\left(\mu_{{\rm obs},i} - v_0\right)^2}{2\left(\sigma_{\rm int}^2 + \sigma_{{\rm obs},i}^2\right)}\right].
\end{equation}

Maximizing this likelihood yields simultaneous estimates of $v_0$ and $\sigma_{\rm int}$ in each bin. As in Sect. \ref{Velocity dispersion modeling and velocity assignment}, the intrinsic velocity dispersion profiles are computed separately for the right ascension ($\mu_{\alpha^*}$) and declination ($\mu_{\delta}$) components, and the total intrinsic dispersion is obtained by combining the two components in quadrature and averaging over them (Eq. \ref{sig_combined}). The statistical uncertainties on these profiles are estimated using bootstrap resampling, also following the procedure outlined in Sect. \ref{Velocity dispersion modeling and velocity assignment}.

\section{Results}
\label{Results}

\begin{figure*}
  \centering
  \includegraphics[width=\textwidth]{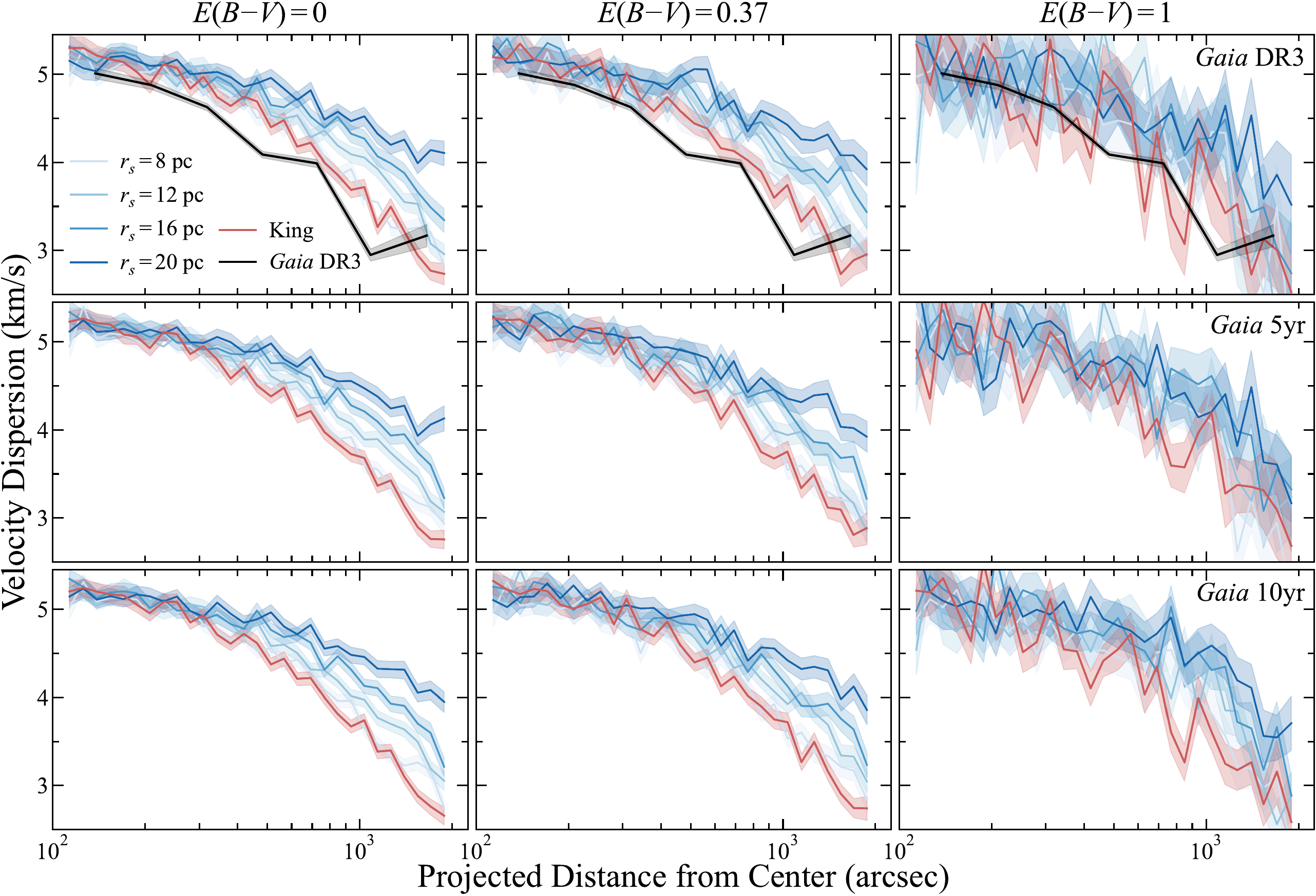}
  \caption{
  Binned velocity dispersions profiles with $1\sigma$ bootstrap uncertainties of the model clusters observed with \textit{Gaia} for three extinction cases. The left column shows $E(B\!-\!V)=0$, the middle column $E(B\!-\!V)=0.37$, and the right column $E(B\!-\!V)=1$. Rows correspond to mission durations of 34 months (DR3), 5 years, and 10 years (top to bottom). The red curve shows the stellar-only model, while the blue curves correspond to the stellar model combined with dark matter halos with different scale radii $r_s$. The black curve shows the binned velocity dispersion profile of M4 derived from \textit{Gaia} DR3 data.}
  \label{fig:vel_disp_gaia}
\end{figure*}

\begin{figure*}
  \centering
  \includegraphics[width=\textwidth]{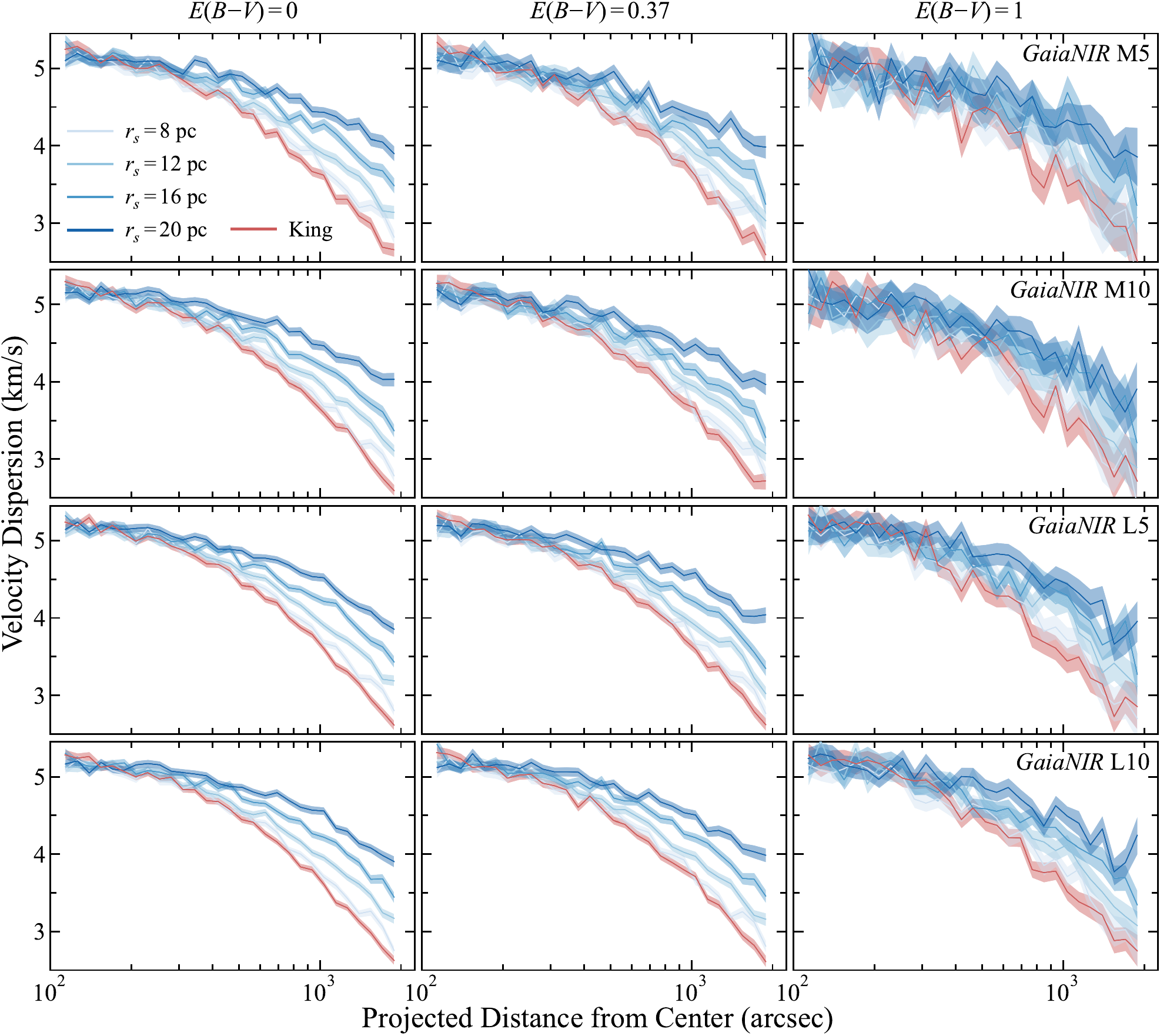}
  \caption{
  Binned velocity dispersions profiles with $1\sigma$ bootstrap uncertainties of the model clusters observed with \textit{GaiaNIR} for three extinction cases. The left column shows $E(B\!-\!V)=0$, the middle column $E(B\!-\!V)=0.37$, and the right column $E(B\!-\!V)=1$. Rows correspond to the M and L mission configuration durations of 5 years, and 10 years. The red curve shows the stellar-only model, while the blue curves correspond to the stellar model combined with dark matter halos with different scale radii $r_s$.  
  }
  \label{fig:vel_disp_gaiarnir}
\end{figure*}

Figures \ref{fig:vel_disp_gaia} and \ref{fig:vel_disp_gaiarnir} present the recovered velocity dispersion profiles for the different model clusters observed by \textit{Gaia}  and \textit{GaiaNIR}, respectively, with $1\sigma$ bootstrap uncertainties. All models use the same stellar distribution generated with \texttt{McLuster}, but the stellar velocities are recalculated for different gravitational potentials: a pure King potential for the stellar-only cluster, and King+NFW potentials for the models including dark matter halos with varying scale radii of $r_s= 8, 12, 16$ and $20~\mathrm{pc}$. For comparison, the intrinsic dispersion profiles, computed before adding observational noise, are shown in Fig. \ref{fig:Velocity_Dispersion_Total_Model}.

Figure \ref{fig:vel_disp_gaia} presents the simulated \textit{Gaia} observations for 34 months (DR3), 5 years ($\sim$DR4), and 10 years ($\sim$DR5). The left column shows the case without extinction, while the middle and right columns show the same models after applying extinctions of $E(B-V)=0.37$ and $E(B-V)=1$, respectively. For DR3 (top panel), the simulated model cluster is compared with real \textit{Gaia} data of M4 to ensure that the simulated values are of the correct order of magnitude. From \textit{Gaia} DR3 we select candidate M4 members by restricting the sample to stars near the cluster center at approximately $\alpha = 245.5^\circ$ and $\delta = -26^\circ$, and by applying a parallax cut of $\varpi = 0-0.72~\mathrm{mas}$ (M4 is located at a distance of $\sim 1.85 ~\mathrm{kpc}$, corresponding to a parallax of $\varpi \sim 0.56~\mathrm{mas}$ \citep{Baumgardt_2021}). The selected parallax range, therefore includes the cluster parallax while allowing for measurement uncertainties, but also includes a substantial number of contaminating stars, particularly background sources, as no lower parallax limit is imposed. To remove these contaminants, we apply a proper-motion cut, since globular clusters typically exhibit distinct systemic motions relative to the field population. Plotting the proper motions of the initially selected stars reveals two separate distributions, where cluster members appear as a group offset from zero proper motion and clearly distinguishable from the bulk of field stars. We therefore select stars within a circular region centered at $(\mu_{\alpha*}, \mu_\delta) = (-12.5, -19)\,\mathrm{mas\,yr^{-1}}$ with a radius of $2\,\mathrm{mas\,yr^{-1}}$. As for the model data, the \textit{Gaia} DR3 data are grouped into 8 logarithmically spaced radial bins between $108$ and $2~160~\mathrm{arcsec}$ ($0.03^\circ$ and $0.6^\circ$), with 3~375 stars in the largest bin and 438 in the smallest bin. The smaller error bars in the observed DR3 data compared to the model error bars rise from differences in binning, with larger bins in the observed data. This comparison with real \textit{Gaia} data of M4 serves as a consistency check rather than a one-to-one model reproduction.

In the absence of extinction, \textit{Gaia} DR3, DR4, and DR5 successfully recover the increase in velocity dispersion at large radii driven by the inclusion of a dark matter halo, also seen in the velocity dispersion profiles of the model cluster before adding measurement noise (see Fig. \ref{fig:Velocity_Dispersion_Total_Model}). Statistical uncertainties decrease with mission duration, improving from a mean value of $\pm 0.13~\mathrm{km~s^{-1}}$ in DR3 to $\pm 0.09~\mathrm{km~s^{-1}}$ in DR5. Halos with scale radii $r_s \leq 8~\mathrm{pc}$ do not produce velocity dispersion enhancements that can be distinguished from the stellar-only model in DR3, DR4, or DR5, and halos with scale radii $r_s = 12~\mathrm{pc}$ produce dispersion enhancements barely distinguishable in DR4 and DR5. For more clearly distinguishable halos ($r_s \geq 16~\mathrm{pc}$), the halo velocity dispersion profiles can be reliably distinguished from the stellar-only profile beyond projected radii of approximately $500-700$ arcsec ($\sim4.5-6.3~\mathrm{pc}$ at M4’s distance). Later data releases (DR4 and DR5) primarily reduce statistical uncertainties but do not extend the radial range over which the halo dispersion profiles can be distinguished from the stellar-only dispersion profile. 

With a moderate extinction of $E(B-V) = 0.37$, the detection significance of dark matter halos relative to the no-extinction case is reduced for \textit{Gaia} data, but the overall trends of the velocity dispersions is still preserved. With $E(B-V) = 1.0$, \textit{Gaia}'s performance degrades severely, and statistical uncertainties increase, from $\pm 0.13~\mathrm{km~s^{-1}}$ in DR3 without extinction to $\pm 0.36~\mathrm{km~s^{-1}}$ in DR3 with an extinction of $E(B-V) = 1.0$. As a result, the recovered velocity dispersion profiles for the dark matter models overlap within $1\sigma$ errors and cannot be distinguished from the stellar-only case, not even for the most extended dark matter halo ($r_s = 20~\mathrm{pc}$). 

Figure \ref{fig:vel_disp_gaiarnir} presents the simulated \textit{GaiaNIR} observations for the model clusters, shown for M5, M10, L5 and L10. The left panels correspond to the extinction free case, and the middle and right panels correspond to the case with $E(B-V)=0.37$ and $E(B-V)=1$, respectively. In the case with no extinction, statistical uncertainties improve from the M5 configuration with a mean value of $\pm 0.10~\mathrm{km~ s^{-1}}$ to the L10 configuration with a mean value of $\pm 0.05~\mathrm{km~s^{-1}}$. All \textit{GaiaNIR} configurations separate the velocity dispersion profiles of dark matter halos with $r_s \geq 12~\mathrm{pc}$ from the stellar-only model. The distinguishable limit of L5 and L10 appears to lie at halo sizes slightly larger than $r_s \simeq 8~\mathrm{pc}$. This represents an improvement over \textit{Gaia} DR5, which requires $r_s \geq 16~\mathrm{pc}$ to achieve a robust detection of a dark matter halo. The radial range over which the halo models become distinguishable decreases to $\sim400-500$ arcsec ($\sim3.6-4.5~\mathrm{pc}$ at the distance of M4), which is approximately 100 arcsec closer to the cluster center than with \textit{Gaia}.

For \textit{GaiaNIR}, an extinction of $E(B-V)=0.37$ retains strong sensitivity to dark matter halos. In all configurations, halos with $r_s \ge 16~\mathrm{pc}$ are robustly detected, and the $r_s=12~\mathrm{pc}$ halo becomes detectable in the M10, L5 and L10 configuration. With $E(B-V) = 1.0$, where \textit{Gaia} loses essentially all dark matter sensitivity, \textit{GaiaNIR} can still recover and, for sufficiently extended halos, distinguish the velocity dispersion profiles from the stellar-only model. In the M configuration, only the most extended halo with $r_s=20~\mathrm{pc}$ can be distinguished. For the L10 configuration, halos with $r_s \geq 16~\mathrm{pc}$ can be distinguished. Statistical uncertainties increase, from $\pm 0.10~\mathrm{km~s^{-1}}$ in M5 without extinction to $\pm 0.23~\mathrm{km~s^{-1}}$ in M5 with an extinction of $E(B-V) = 1.0$.

\section{Discussion}
\label{Discussion}

The results demonstrate that the detectability of dark matter halos in globular clusters depends strongly on both observational conditions, such as extinction and instrumental sensitivity, and on the intrinsic properties of the halo. In Sect. \ref{Detectability}, we interpret the mock velocity dispersion profiles and assess the ability of the two missions to distinguish the presence of dark matter halos. In Sect. \ref{Limitations}, we discuss the main assumptions of the modeling, and in Sect. \ref{FutureWork} we outline directions for future work.  

\subsection{Detectability of dark matter halos with \textit{Gaia} and \textit{GaiaNIR}}
\label{Detectability}

For both \textit{Gaia} and \textit{GaiaNIR}, the sensitivity to distinguish dark matter halos depends primarily on the halo mass and spatial extent. When observed with no extinction, we find that \textit{Gaia} is able to distinguish the most extended dark matter halos in the M4-like model with $r_s \geq 16 ~\mathrm{pc}$, from the stellar-only model, with a sensitivity that improves with longer mission durations (Fig. \ref{fig:vel_disp_gaia}, left panel). When observed with moderate extinction ($E(B-V) = 0.37$), typical of the outer disk and similar to the measured reddening of M4 \citep{reddening_M4}, the detection significance of dark matter halos is reduced (Fig. \ref{fig:vel_disp_gaia}, middle panel). With high extinction ($E(B-V) = 1$), such as the extinction present in the Galactic midplane and bulge, \textit{Gaia} loses all sensitivity to dark matter halos (Fig. \ref{fig:vel_disp_gaia}, right panel). This highlights that the primary limitation of \textit{Gaia} in resolving dark matter halos occurs in dust-obscured environments, where its restriction to visible wavelengths significantly reduces its effectiveness.

\textit{GaiaNIR} shows improvements over \textit{Gaia} across all tested configurations (Fig. \ref{fig:vel_disp_gaiarnir}). In both the cases with and without extinction, the statistical uncertainties are reduced, enabling stronger constraints on the presence of dark matter halos. This improvement is most evident under high extinction ($E(B-V) = 1$), where \textit{GaiaNIR} maintains the ability to distinguish extended halos with $r_s = 20~\mathrm{pc}$, and, with the L10 configuration, it can even detect halos with $r_s \geq 16~\mathrm{pc}$. 

Between the two \textit{GaiaNIR} telescope configurations, the L design (3.5 m aperture) outperforms the M design (1.7 m aperture). Since the collecting area scales as  $D\times H$ while $H$ is fixed to 0.5~m, the L configuration gathers twice as many photons than the M configuration. This corresponds a factor of $\sqrt{2}$ improvement in the astrometric precision at fixed magnitude, leading to enhanced detectability of dark matter halos.

The astrometric improvement of \textit{GaiaNIR} over \textit{Gaia} is particularly pronounced for faint stars, where the uncertainties of \textit{Gaia} increase rapidly due to the growing importance of readout noise, while the lower effective readout noise of the APD detectors in \textit{GaiaNIR} leads to a much slower degradation in precision (see Sect.~\ref{Instrumental performance and astrometric errors} and Fig.~\ref{fig:Parallax_error_GaiaNIR}). As a result, \textit{GaiaNIR} retains significantly better precision for faint stars, increasing the number of stars that can be used for kinematic measurements.

In practice, \textit{GaiaNIR} observations would also be combined with existing \textit{Gaia} data, or data from other missions, e.g. Roman \citep{spergel2015widefieldinfrarredsurveytelescopeastrophysics}, extending the astrometric time baseline and improving the overall precision of the combined dataset. In regions of low extinction, this leads to improved precision for stars already observed by \textit{Gaia}, while in regions of high extinction \textit{GaiaNIR} enables observations of stars that are not accessible in the visible. Thereby \textit{GaiaNIR} will increase both the quality and the number of measurements especially if combined with other NIR data sets.

Overall, these results demonstrate that \textit{GaiaNIR} would enable the study of globular cluster dynamics in regions of high extinction that are currently inaccessible to \textit{Gaia}, such as the Galactic bulge and midplane. This expanded coverage would increase the number of clusters available for dynamical studies, allowing for better statistical constraints on dark matter in globular clusters and stronger tests of formation scenarios that predict dark matter retention.

\subsection{Modeling assumptions}
\label{Limitations}

The present analysis provides a comparative assessment of the ability of \textit{Gaia} and \textit{GaiaNIR} to constrain cluster kinematics. The analysis is based on a number of modeling assumptions, which are outlined below. 

In this work, we do not consider joint astrometric solutions combining \textit{Gaia} and \textit{GaiaNIR}, as the number of common stars is uncertain. Instead, the analysis is based on standalone \textit{GaiaNIR} astrometric uncertainties derived from a physically motivated instrument model that combines stellar spectra with assumptions on detector performance and photon statistics (see Sect.~\ref{Instrumental performance and astrometric errors}). These estimates depend on current design assumptions, as the final instrument characteristics have not yet been fully defined, and should therefore be regarded as indicative (Hobbs et al. in prep.). Nevertheless, this approach provides a framework for simulating \textit{GaiaNIR} observations and enables exploration of its expected performance across a wide range of astrophysical applications also beyond the specific case considered here \citep[e.g.][Jiménez-Arranz et al. in prep.]{Marie}.

A number of observational effects are not included in the modeling. In particular, we do not account for crowding, which becomes important in the central regions of the cluster. In such dense environments as globular clusters, crowding leads to unresolved sources and increased astrometric uncertainties, reducing the number of reliable stellar measurements \citep[e.g.][]{Crowding_Gaia_2017, Gaia_extinction_2023}. This limitation will also be especially relevant for \textit{GaiaNIR}, as it will probe densely populated regions of the central Galaxy. We also do not account for spatially correlated astrometric systematics, which can introduce position-dependent offsets in proper motions and bias measurements of the internal kinematics \citep{Vasiliev_2019}. In addition, the model cluster contains only genuine members, and thus does not include foreground or background field-star contamination, which in real observations would also affect the observed kinematics. However, improved astrometric precision, as expected from \textit{GaiaNIR}, will enhance the ability to distinguish cluster members from field stars, providing a further improvement over what can be achieved with \textit{Gaia} alone.

Various intrinsic astrophysical effects are also simplified or not included in our model. In particular, unresolved binary systems are not included. Such systems can bias velocity measurements, as they act as more massive kinematic tracers than single stars \citep[e.g.][]{Bianchini_2016_binaries}. However, the binary fraction in globular clusters is typically low compared to the Galactic field \citep[e.g.][]{2012_binaries}. Mass segregation is also not modeled. In real globular clusters, more massive stars tend to sink toward the center over time, while lower-mass stars populate the outer regions \citep[e.g.][]{Teodori_2024_mass_segregation}. Because more massive stars typically exhibit lower velocity dispersions than lower-mass stars, this effect can influence the velocity dispersion profile. In the present models, however, the velocity dispersion profiles are rescaled to match at the center, so any central deviations are largely absorbed by this normalization. Furthermore, tidal interactions with the Galactic potential are not included. Such interactions could affect the outer kinematics, for example through tidal heating or the stripping of stars \citep{Lane_2010}. These effects are expected to be most important for clusters located in regions where the Galactic tidal field is strong, such as the Galactic bulge or disk.

Our kinematic analysis also involves methodological simplifications. The velocity dispersion profiles are derived under the assumption of velocity isotropy in the Jeans equation (Eq. \ref{Jeans}). While this is well supported observationally in the central regions of globular clusters \citep{Jindal_2019_isotropy}, it is less certain in the outer regions where the dark matter signal emerges. The velocity dispersion profiles are also derived using radial binning, which introduces a dependence on bin size. If the goal were to estimate precise halo detection thresholds for the two missions rather than perform a comparative analysis, methods that avoid binning would be preferable to reduce potential information loss. In addition, we compute the total projected velocity dispersion by combining the $\sigma_{\mu_\alpha*}$ and $\sigma_{\mu_\delta}$ components, rather than analyzing them separately. While this maximizes statistical precision, it does not account for possible correlations between the two components, which are provided in the \textit{Gaia} astrometric covariance matrix \citep[e.g.][]{Everall_2021_COV}.

Finally, it should be noted that another way to detect the presence of dark matter in clusters is through a discrepancy between the dynamical mass, derived from the velocity dispersion, and the stellar mass estimated from star counts. This approach is commonly applied in systems such as ultra-faint dwarf galaxies, where the dynamical mass significantly exceeds the stellar mass \citep[e.g.][]{Simon_2019}. However, in our models, the central rescaling of the velocity dispersions, removes the constraint on the total mass implied by the velocity dispersion. As a result, this method does not provide a reliable measurement of the dark matter content, and the focus should instead be on the radial variation of the velocity dispersion profile.

While the assumptions and limitations discussed above affect the observed velocity dispersion and thereby the detection thresholds on dark matter halos, it is important to note that since both \textit{Gaia} and \textit{GaiaNIR} are analyzed under identical modeling assumptions, the relative comparison between the two missions is expected to be relatively robust to the simplifications made in this work, even if the precise detection thresholds may shift when more realistic modeling is applied.

\subsection{Future work}
\label{FutureWork}

We plan to extend this analysis to a broader range of cluster environments, which will be essential for fully assessing the scientific potential of \textit{GaiaNIR}. In particular, modeling globular clusters beyond M4, and placing them in different Galactic environments such as the bulge and disk, will allow for a more comprehensive evaluation under varying levels of extinction and stellar density. In these regions, we plan to include a treatment of crowding effects, which would reduce the number of usable stars and degrade astrometric precision, but provide more realistic performance estimates. In addition, we aim to extend the analysis to younger embedded and open clusters, which host warmer and more massive stars. This will allow us to evaluate \textit{GaiaNIR}'s performance across a broader range of stellar populations, and provide a even more complete picture of its capabilities.

\section{Conclusions}
\label{conlusions}

The \textit{Gaia} mission has transformed globular cluster research by delivering precise astrometric and photometric data, allowing accurate measurements of memberships, distances, and motions across much of the MW’s cluster population. However, interstellar extinction still limits observations, especially for clusters in the Galactic bulge and inner disk, where dust reduces data completeness and precision. While halo clusters are less affected, studying obscured regions is essential for understanding the MW’s inner structure and evolution. The proposed \textit{GaiaNIR} mission addresses this by extending observations into the NIR, where dust effects are significantly weaker. By providing more complete and precise astrometric measurements, \textit{GaiaNIR} will place stronger constraints on the internal dynamics of globular clusters and enhance our ability to detect potential dark matter halos.

In this work, we assess how well both  \textit{Gaia} and \textit{GaiaNIR} can distinguish dark matter halos of different sizes around globular clusters under varying extinction conditions. To do so, we construct a model globular cluster with properties similar to those of M4 and modify the stellar velocities to follow either a stellar-only density distribution or a stellar component embedded in dark matter halos of different sizes. Observational noise corresponding to \textit{Gaia} and \textit{GaiaNIR} is then applied to generate mock observations for each mission. The main results of the paper can be summarized as follow:
\begin{enumerate}

\item \textit{Gaia} can resolve extended dark matter halos when the cluster is observed with no extinction (Fig. \ref{fig:vel_disp_gaia}, left panels) and with a moderate extinction of $E(B-V)=0.37$ (Fig. \ref{fig:vel_disp_gaia}, middle panels), but with lower significance.

\item \textit{Gaia} fails to distinguish halos when the cluster is observed with high extinction of $E(B-V)=1$ (Fig. \ref{fig:vel_disp_gaia}, right panels).

\item \textit{GaiaNIR} reduces the statistical uncertainties when the cluster is observed both with and without extinction (Fig. \ref{fig:vel_disp_gaiarnir}).

\item \textit{GaiaNIR} is able to resolve extended halos even under high-extinction conditions of $E(B-V)=1$ (Fig. \ref{fig:vel_disp_gaiarnir}, right panels).

\end{enumerate}
Overall, these results indicate that \textit{GaiaNIR} would provide tighter constraints on the kinematic signatures detected with \textit{Gaia} and enable observations of globular clusters in regions of high extinction that are currently inaccessible to \textit{Gaia}.

\section*{Data availability}

Both \textit{Gaia} and \textit{GaiaNIR} mock catalogues may be obtained from the corresponding author upon reasonable request.

\begin{acknowledgements}

IH acknowledges funding by the European Union under the Horizon Europe Marie Skłodowska-Curie Actions Doctoral Network grant agreement no. 101072454 @HorizonEU research and innovation programme.

DH and OJA acknowledges funding from ``Swedish National Space Agency 2023-00154 David Hobbs The GaiaNIR Mission'' and ``Swedish National Space Agency 2023-00137 David Hobbs The Extended Gaia Mission''.

This work has made use of data from the European Space Agency (ESA) mission {\it Gaia} (\url{https://www.cosmos.esa.int/gaia}), processed by the {\it Gaia} Data Processing and Analysis Consortium (DPAC, \url{https://www.cosmos.esa.int/web/gaia/dpac/consortium}). Funding for the DPAC has been provided by national institutions, in particular the institutions participating in the {\it Gaia} Multilateral Agreement.

\\
\textit{Software:}
\textsc{McLuster} \citep{McLuster},
\textsc{Pickles Atlas} \citep{Pickles_Atlas},
\textsc{ipython} \citep{ipython}, 
\textsc{jupyter} \citep{jupyter},
\textsc{matplotlib} \citep{matplotlib},
\textsc{numpy} \citep{numpy},
\textsc{pandas} \citep{pandas...paper, pandas...software},
\textsc{pygaia} \citep{pygaia}

\end{acknowledgements}

\bibliography{citations}
\bibliographystyle{aa} 

\clearpage

\begin{appendix}

\section{Empirical relations}
\label{Empirical relations}

Polynomial transformations between the Johnson-Cousins and \textit{Gaia} EDR3 photometric systems \citep{Riello_2021}.

\begin{equation}
\begin{aligned}
G - V = & \; 0.04749 - 0.0124(B - V) \\
      & - 0.2901(B - V)^2 + 0.02008(B - V)^3
\end{aligned}
\label{G_1}
\end{equation}

\begin{equation}
\begin{aligned}
G - V = & -0.01597 - 0.02809(V - I_c) - 0.2483(V - I_c)^2  \\
      & + 0.03656(V - I_c)^3 - 0.002939(V - I_c)^4
\end{aligned}
\label{G_2}
\end{equation}

\begin{equation}
\begin{aligned}
G_{\text{BP}} - G_{\text{RP}} = & -0.03298 + 1.259 (V - I_c) -0.1279 (V - I_c)^2 \\
& + 0.01631 (V - I_c)^3
\end{aligned}
\end{equation}

Polynomial transformations between the 2MASS and \textit{Gaia} EDR3 photometric systems \citep{Riello_2021}.

\begin{equation}
G- H = -0.1048 + 2.011 (G_{\text{BP}} - G_{\text{RP}}) - 0.1758 (G_{\text{BP}} - G_{\text{RP}})^2
\label{$G$-H}
\end{equation}

\begin{equation}
G- J = 0.01798 + 1.389 (G_{\text{BP}} - G_{\text{RP}}) -0.09338 (G_{\text{BP}} - G_{\text{RP}})^2
\end{equation}

\begin{equation}
\begin{aligned}
G- K_s = & -0.0981 + 2.089 (G_{\text{BP}} - G_{\text{RP}}) \\
& -0.1579 (G_{\text{BP}} - G_{\text{RP}})^2
\end{aligned}
\end{equation}
\\

\section{King and NFW models}
\label{King and NFW profiles}

For the cluster without a dark matter halo, we use a King model with the density distribution:

\begin{equation}
\rho\left(\Psi \right) = \rho_1 \left[ e^{\Psi / \sigma^2} \operatorname{erf} \left( \frac{\sqrt{\Psi}}{\sigma} \right) 
- \sqrt{\frac{4\Psi}{\pi\sigma^2}} \left( 1 + \frac{2\Psi}{3\sigma^2} \right) \right]
\end{equation}

Where erf is the error function, $\Psi\equiv-\Phi+\Phi_0$ is the relative potential, $\Phi$ is the gravitational potential, and $\Phi_0$ is some constant. $\rho_1$ is the central density, assumed to be $\rho_1 = 2~\mathrm{M}_{\odot}~pc^{-3}$, and $\sigma$ is the constant velocity dispersion, assumed to be $\sigma = 3.6~\mathrm{km~s^{-1}}$. These parameters are chosen to match the \texttt{McLuster} output and yield a total cluster mass of approximately $1.5 \times 10^5~\mathrm{M}_{\odot}$ within the tidal radius.

To determine the density and potential as functions of radius, we solve Poisson's equation:

\begin{align}
\frac{d}{dr} \left( r^2 \frac{d\Psi}{dr} \right) &= 
-4\pi G\rho_1 r^2 \left[ e^{\Psi / \sigma^2} \operatorname{erf} \left( \frac{\sqrt{\Psi}}{\sigma} \right) 
- \sqrt{\frac{4\Psi}{\pi\sigma^2}} \left( 1 + \frac{2\Psi}{3\sigma^2} \right) \right]
\label{poisson}
\end{align}

We assume that $\Psi(0) = W_0 \sigma^2$, where $W_0=7$ is the central dimensionless potential, and  that $\frac{d\Phi}{dr} = 0$ at $r=0$, due to spherical symmetry.

From the density profile, the mass profile follows:

\begin{equation}
M(r) = 4\pi \int_0^r \rho(r') r'^2 ~dr'
\label{mass}
\end{equation}

\begin{figure*} 
    \centering
    \includegraphics[width=\textwidth]{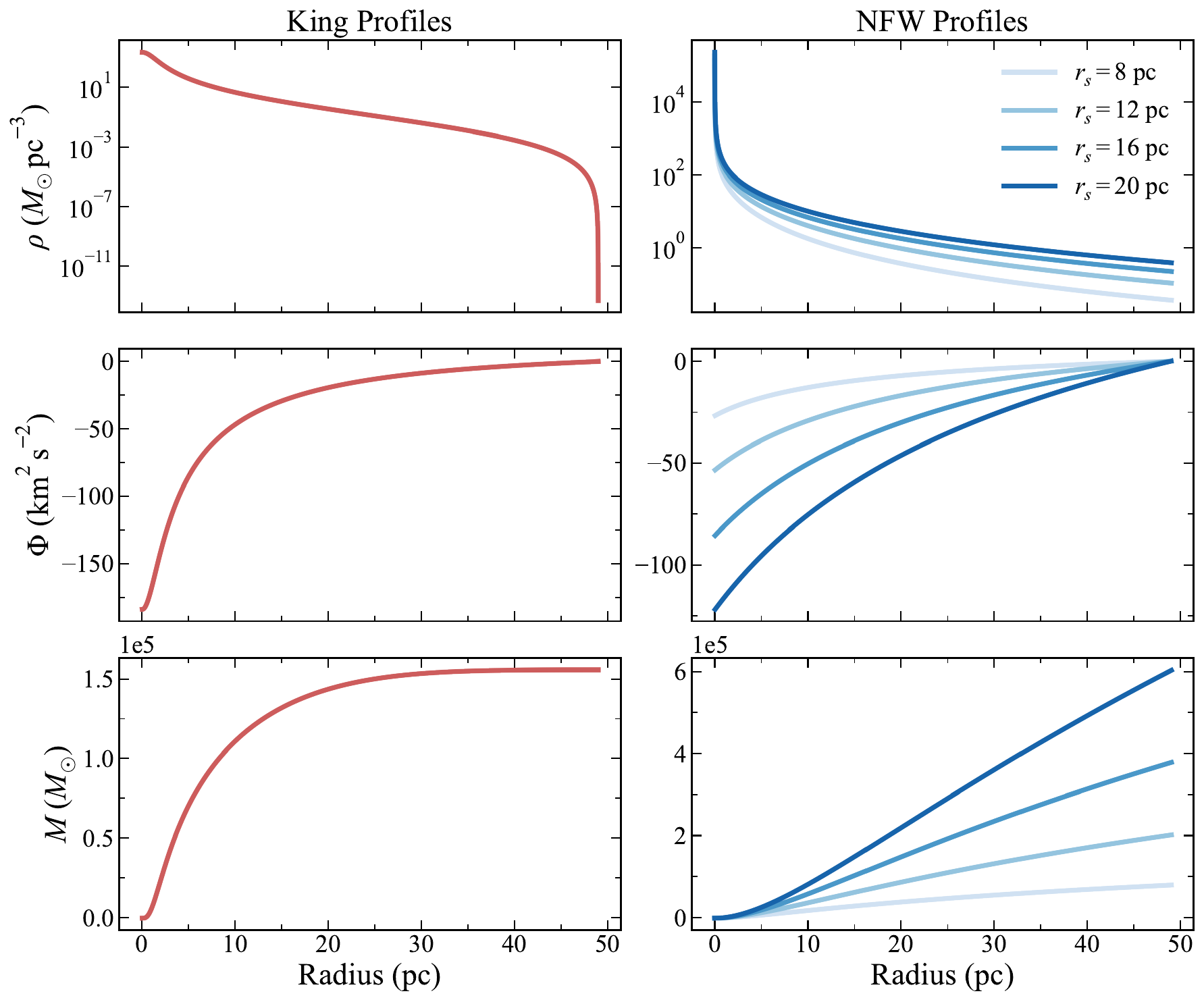}
    \caption{Left panel: The density, potential, and mass profiles of the King model as functions of radius. Right panel: The density, potential, and mass profiles of the NFW model as functions of radius}
    \label{fig:King_vs_NFW_Profiles}
\end{figure*}

The density profile ($\rho$), gravitational profile ($\Phi$) and mass ($M$) profile of the king model as function of radius are illustrated in Fig.\ref{fig:King_vs_NFW_Profiles} (left panel).

To model the cluster embedded in a dark matter halo, we include an additional NFW model in the gravitational potential.

The gravitational potential of an NFW model is given by:

\begin{equation}
\Phi(r) = - 4\pi G\rho_0 r_s^2 \frac{\ln(1 + r/r_s)}{r}
\end{equation}

Where $r_s$ is the scale radius, which we vary between $8$, $12$, $16$, and $20~\mathrm{pc}$. $\rho_0$ is the characteristic density, chosen to be $\rho_0 = 11~\mathrm{M_\odot pc^{-3}}$, to produce dark matter masses comparable to the stellar mass, and $r_s$ is the scale radius. 

The density profile of the NFW halo is given gy:

\begin{equation}
    \rho(r) = \frac{\rho_0}{\frac{r}{r_s} \left(1 + \frac{r}{r_s} \right)^2}
\label{NFW_density}
\end{equation}

Using Eq.~\ref{mass}, the corresponding mass profile becomes

\begin{equation}
M(r) = 4\pi \rho_0 r_s^3 \left[ \ln \left(1 + \frac{r}{r_s} \right) - \frac{r}{r_s + r} \right]
\end{equation}

The density profile ($\rho$), gravitational profile ($\Phi$) and mass ($M$) profile of the NFW model as functions of radius are shown in Figure \ref{fig:King_vs_NFW_Profiles} (right panel).

\section{Interstellar extinction}
\label{Ap:extinction}
To add extinction to the stellar spectra, we adopt the fit to Seaton’s extinction law \citep{Seaton} given by \citet{Reddening}:

\begin{align}
\frac{E(\lambda - V)}{E(B - V)} =\
& C_1 + C_2 \lambda^{-1} + \frac{C_3}{\left[\lambda^{-1} - \left(\frac{\lambda_0^{-1}}{\lambda^{-1}}\right)^2\right] + \gamma^2} \notag\\
    & + C_4 \left[ 0.539(\lambda^{-1} - 5.9)^2 + 0.0564(\lambda^{-1} - 5.9)^3 \right]
\end{align}

where $\lambda$ is in $\mu\mathrm{m}$, $\lambda_0^{-1} = 4.595~\mu\mathrm{m}^{-1}$, $\gamma = 1.051~\mu\mathrm{m}^{-1}$, $C_1 = -0.38$, $C_2 = 0.74~\mu\mathrm{m}$, $C_3 = 3.96\ \mu\mathrm{m}^{-2}$, and $C_4 = 0.26\ \mu\mathrm{m}^{-1}$ (or $C_4 = 0$ if $\lambda^{-1} > 5.9$).

This law gives the wavelength-dependent color excess $E(\lambda - V)$, i.e. the additional extinction at wavelength $\lambda$ relative to the $V$ band. It is expressed in units of the standard color excess $E(B - V)$ and can be written as:

\begin{equation}
E(\lambda - V) = k(\lambda) \times E(B - V),
\end{equation}

where $k(\lambda)$ is the extinction curve. The total extinction $A(\lambda)$ is related to $E(\lambda - V)$ through:

\begin{equation}
E(\lambda - V) = A(\lambda) - A(V),
\end{equation}

with $A(V) = R_V \times E(B - V)$, where $R_V$ is the total-to-selective extinction ratio. We adopt $R_V = 3.1$, the typical MW value \citep{Szomoru_1999}.

Substituting gives:

\begin{equation}
A(\lambda) = \left[ R_V + k(\lambda) \right] \times E(B - V),
\end{equation}

which allows the extinction at each wavelength to be calculated directly from a given color excess and the extinction curve.

To apply this extinction to the stellar spectra, we relate the magnitude difference caused by extinction to the corresponding change in flux, using Pogson’s law. In the case of extinction, $F_1$ corresponds to the intrinsic (unextinguished) flux $F_{\lambda,\mathrm{intrinsic}}$, and $F_2$ corresponds to the observed (extinguished) flux $F_{\lambda,\mathrm{observed}}$. The magnitude difference is then exactly the extinction at wavelength $\lambda$:

\begin{equation}
A(\lambda) = m_{\lambda, \mathrm{observed}} - m_{\lambda, \mathrm{intrinsic}}
\end{equation}

Substituting into Pogson’s law and multiplying by $F_{\lambda,\mathrm{intrinsic}}$ gives the spectra affected by extinction:

\begin{equation}
F_{\lambda,\mathrm{observed}} = F_{\lambda,\mathrm{intrinsic}} \times 10^{-0.4 A(\lambda)}
\end{equation}

\end{appendix}

\end{document}